\pdfoutput=1
\documentclass[twocolumn,english,sort]{scrartcl}
\usepackage{helvet}

\usepackage[T1]{fontenc}
\usepackage[latin9]{inputenc}
\usepackage[letterpaper]{geometry}
\usepackage{babel}
\usepackage{units}
\usepackage{amsmath}
\usepackage{graphicx}
\usepackage{setspace}
\usepackage{esint}
\usepackage[unicode=true,pdfusetitle,
 bookmarks=true,bookmarksnumbered=false,bookmarksopen=false,
 breaklinks=false,pdfborder={0 0 0},backref=false,colorlinks=false]
 {hyperref}

\makeatletter

\providecommand{\tabularnewline}{\\}

\numberwithin{equation}{section}
\numberwithin{figure}{section}

\usepackage[comma,super]{natbib}
\renewcommand\@biblabel[1]{$^{#1}$}
\usepackage[modulo]{lineno}
\date{}
\usepackage{remreset}

\DeclareMathSizes{10}{10}{10}{10}

\makeatother

\begin{document}

\title{K-edge subtraction vs. A-space processing for x-ray imaging of contrast
agents: SNR}

\author{Robert E.  Alvarez\thanks{ralvarez@aprendtech.com}\thanks{Code for figures available online\cite{alvarez_kedge_paper_code1}}}
\maketitle
\begin{abstract}
Purpose: To compare two methods that use x-ray spectral information
to image externally administered contrast agents: K-edge subtraction
and basis-function decomposition (the A-space method),

Methods: The K-edge method uses narrow band x-ray spectra with energies
infinitesimally below and above the contrast material K-edge energy.
The A-space method uses a broad spectrum x-ray tube source and measures
the transmitted spectrum with photon counting detectors with pulse
height analysis. The methods are compared by their signal to noise
ratio (SNR) divided by the patient dose for an imaging task to decide
whether contrast material is present in a soft tissue background.
The performance with iodine or gadolinium containing contrast material
is evaluated as a function of object thickness and the x-ray tube
voltage of the A-space method. 

Results: For a tube voltages above 60 kV and soft tissue thicknesses
from 5 to 25 g/cm\textasciicircum{}2, the A-space method has a larger
SNR per dose than the K-edge subtraction method for either iodine
or gadolinium containing contrast agent. 

Conclusion: Even with the unrealistic spectra assumed for the K-edge
method, the A-space method has a substantially larger SNR per patient
dose.

\vspace{0.2cm}

\hspace{-0.1cm}Key Words: spectral x-ray, K-edge subtraction, basis
decomposition, contrast agent, photon counting,
\end{abstract}

\section{INTRODUCTION}

K-edge subtraction was one of the first methods to use x-ray spectral
information to improve the visibility of contrast agents injected
into the body\cite{jacobson_dichromatic_1953,jacobson_edholm_iodine_1959,riederer_selective_iodine_Kedge_1977}.
An alternative method utilizing spectral information is the basis
decomposition method\cite{Alvarez1976} (the A-space method). Alvarez\cite{Alvarez2010}
showed this method can be used to provide near optimal signal to noise
ratio (SNR) \cite{Tapiovaara1985} with low energy-resolution measurements.
Although both methods use spectral information, they are quite different
and an interesting question is which one provides a better SNR per
dose for detecting an externally adminstered contrast agent in a soft
tissue background?

This paper examines that question for idealized spectra and detectors.
For K-edge subtraction, monoenergetic spectra with energies just below
and above the K-edge energy and a quantum noise-limited, negligible
pileup photon counting detector are used. For the A-space method,
a broad spectrum x-ray tube source is used with an ideal photon counting
detector with pulse height analysis (PHA). 

The imaging task is to detect contrast material embedded in soft tissue.
The signal to noise ratios of the two methods are compared for equal
dose, which is approximated as the absorbed energy. Since, in general,
the square of the SNR is proportional to dose, the parameter compared
is $\nicefrac{SNR^{2}}{Dose}$. This parameter is computed as a function
of soft-tissue object thickness and the x-ray tube voltage for contrast
agents containing iodine or gadolinium.

The imaging task does not measure the full capability of either method.
The A-space method extracts a great deal more information about the
object than simply the presence of the contrast agent\cite{Alvarez,Alvarez1982,Lehmann1981}.
The K-edge subtraction method can discriminate against body materials
with different compositions so long as their attenuation coefficient
is continuous at the K-edge energy. Nevertheless the imaging task
used allows us to directly compare the fundamental performance of
the two methods for an important clinical application. More general
comparisons including the presence of other body materials and practical
limitations on the x-ray source spectra for the K-edge subtraction
method are under research. 

Recent examples of work in this area include Roessl and Proksa\cite{Roessl2007},
who applied the A-space method to image contrast agents as did Zimmerman
and Schmidt\cite{ZimmermanSPIEMedImg3Destimators2016}. Dilmanian
et al.\cite{dilmanian_synchrotron_Kedge1997} used a synchrotron radiation
source at a nuclear physics laboratory to image contrast agents with
K-edge subtraction. Shikhaliev\cite{shikhaliev2012photonCountKedgeFiltered}
pre-filtered a broad spectrum source with high atomic number materials
to provide a bi-modal spectrum. The filtered transmitted spectrum
was measured with a photon counting detector with PHA and the SNR
of the contrast material thickness was computed. None of these papers
compared the K-edge subtraction method to the A-space method directly.

\section{METHODS}

In this section, the imaging task for the signal to noise ratio definition
is described. Then expressions for the SNR and absorbed energy with
the K-edge subtraction and A-space methods are derived. Finally, the
SNR per absorbed energy is computed as a function of the object thickness
and the tube voltage for contrast agents containing iodine or gadolinium.

\subsection{The imaging task\label{sub:The-imaging-task}}

\begin{figure}
\centering{}\includegraphics[scale=0.55]{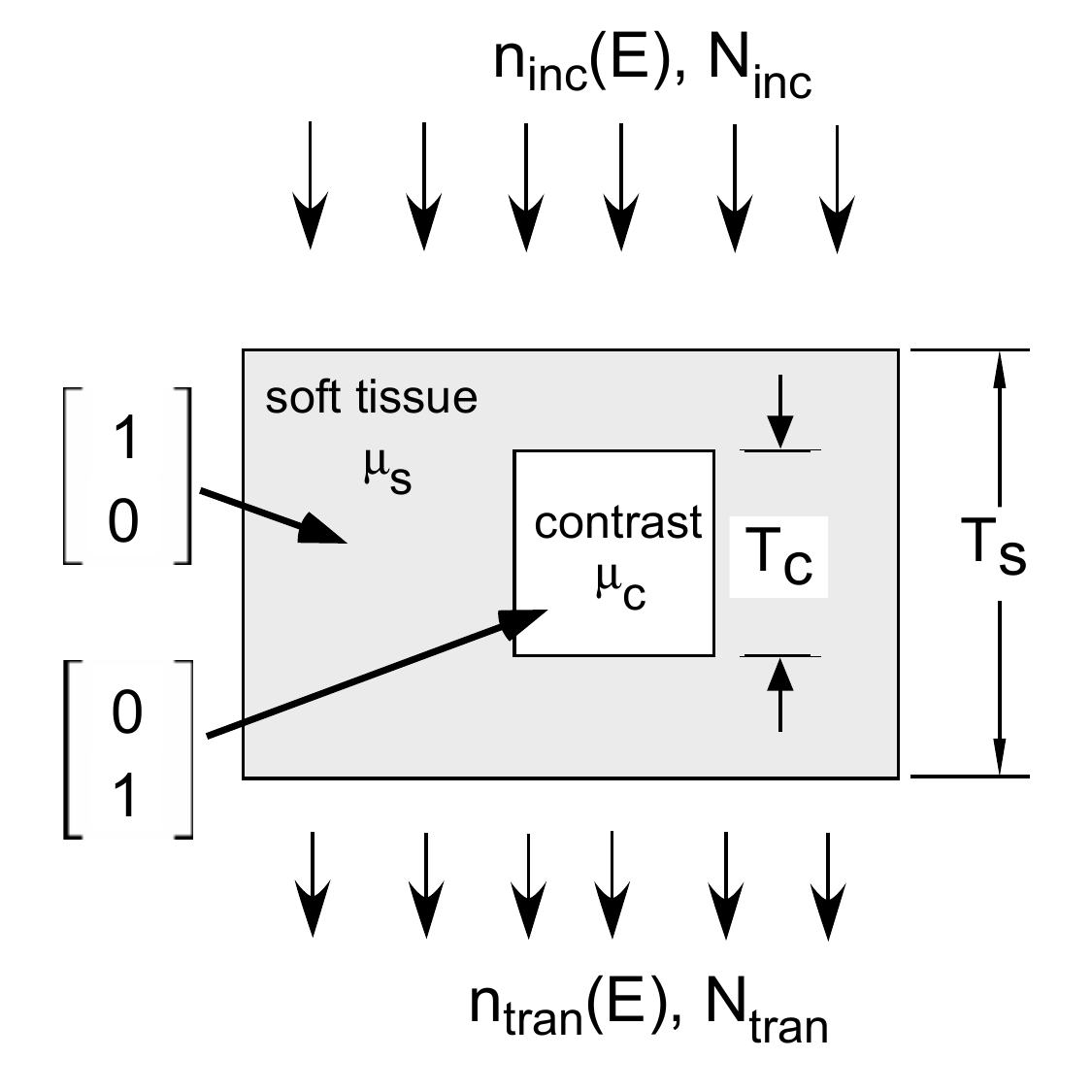}

\protect\caption{The imaging task is to detect the presence of an externally administered
contrast agent from measurements of the transmitted energy spectrum,
$n_{tran}(E)$, where $E$ is the x-ray energy. The incident spectrum
is $n_{inc}(E)$ and the total number of incident photons is $N_{inc}=\int n_{inc}(E)dE$.
The soft tissue material has attenuation coefficient $\mu_{s}(E)$
and thickness $T_{s}$, while the contrast material has attenuation
coefficient $\mu_{c}(E)$ and thickness $T_{c}$. For A-space processing,
a basis set consisting of the background and contrast materials' attenuation
coefficients is used. With this basis set, the coefficients, the $\mathbf{a}$
vectors, for the soft tissue is $\left[0,1\right]^{T}$ and the contrast
material $\left[1,0\right]^{T}$ as shown in the figure. Column vectors
are used and the symbol $T$ denotes the transpose. \label{fig:Imaging-task}}
\end{figure}

The imaging task assumes the object shown in Fig. \ref{fig:Imaging-task}.
The task decides whether a contrast material is present from measurements
of the transmitted x-ray energy spectrum.

Assuming normally distributed noise, the probability of error depends
only on the signal to noise ratio\cite{VanTrees2013}, where the signal
is the square of the difference in the expected values of the measurements
between the soft tissue only and the soft tissue plus contrast materials
and the noise is the variance of the measurements. The contrast material
thickness will be assumed to be sufficiently small that the variance
is essentially the same in the soft tissue-only and the contrast regions.

\subsection{Ideal K-edge subtraction\label{sub:Ideal-K-edge-subtraction}}

\begin{figure}
\centering{}\includegraphics[scale=0.65]{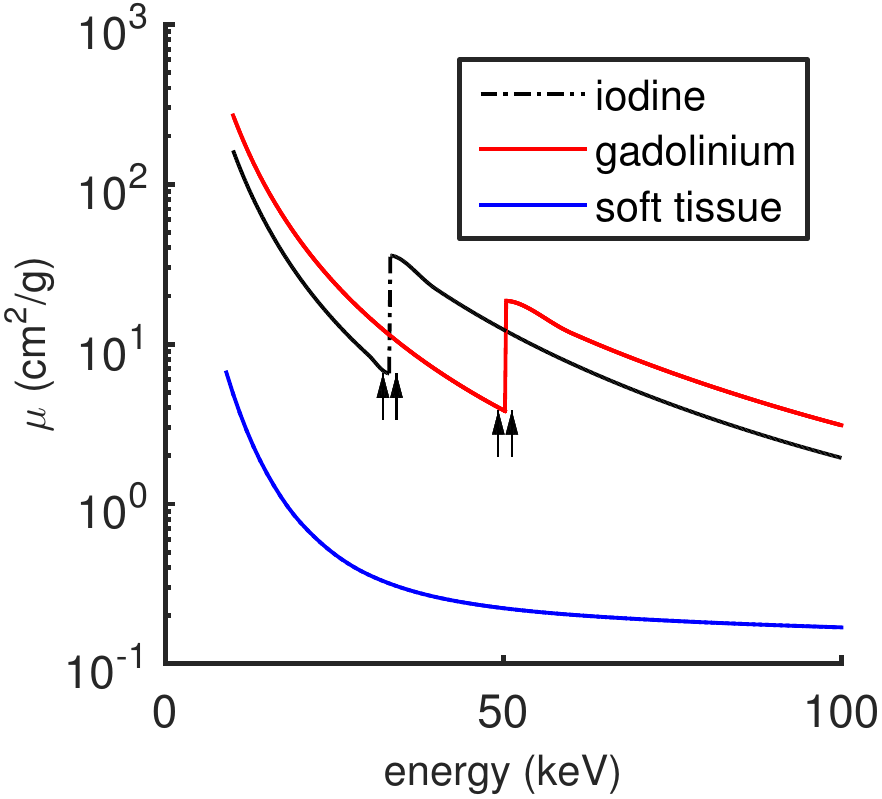} 

\begin{tabular}{|l|c|c|c|}
\hline 
 & \textbf{}%
\begin{tabular}{c}
\textbf{$E_{K}$ }\tabularnewline
$\left(keV\right)$\tabularnewline
\end{tabular} & \textbf{}%
\begin{tabular}{c}
\textbf{$\mu(E_{K-})$}\tabularnewline
($cm^{2}/g)$\tabularnewline
\end{tabular} & \textbf{}%
\begin{tabular}{c}
\textbf{$\mu(E_{K+})$}\tabularnewline
($cm^{2}/g)$\tabularnewline
\end{tabular}\tabularnewline
\hline 
\textbf{iodine} & 33.17 & 10.18 & 31.59\tabularnewline
\hline 
\textbf{gadolinium} & 50.24 & 4.88 & 14.55\tabularnewline
\hline 
\end{tabular}

\protect\caption{X-ray attenuation coefficients of the high atomic number elements
used in contrast agents as a function of x-ray energy. The attenuation
coefficient of soft tissue is also plotted for comparison. The table
shows the K-edge energy and the attenuation coefficients just below
and above it for iodine and gadolinium.\label{fig:Mu-contrast-vs-egy}}
\end{figure}

Figure \ref{fig:Mu-contrast-vs-egy} shows the attenuation coefficients
of iodine and gadolinium as well as soft tissue. Notice the sharp
discontinuities of the coefficients of the contrast materials. The
absorption edge energies are unique to each element. The energies
and the attenuation coefficients just above and below the discontinuities
are shown in the table in the figure. Soft tissue and other biological
materials have attenuation coefficients that are continuous throughout
the diagnostic energy range.

For the ideal K-edge method, we use delta function x-ray spectra with
energies just below and just above the K-edge energy, $E_{k-}$and
$E_{k+}$, and compute the difference of the logarithm of the number
of transmitted photons. Referring to Fig. \ref{fig:Imaging-task},
the expected values of the transmitted photon counts with the two
spectra are $N_{K-}$ and $N_{K+}$
\begin{equation}
\begin{array}{ccc}
N_{K-} & = & \left(\nicefrac{N_{0}}{2}\right)e^{-\mu_{s}\left(E_{k-}\right)(t_{s}-t_{c})-\mu_{c}\left(E_{k-}\right)t_{c}}\\
N_{K+} & = & \left(\nicefrac{N_{0}}{2}\right)e^{-\mu_{s}\left(E_{k+}\right)(t_{s}-t_{c})-\mu_{c}\left(E_{k+}\right)t_{c}}
\end{array}.\label{eq:nKs}
\end{equation}

\noindent{}In this equation, $N_{0}$ is the sum of the incident
photons of both spectra, $\mu_{s}(E)$ is the soft-tissue attenuation
coefficient at x-ray energy $E$, $t_{s}$ is the soft tissue material
thickness, $\mu_{c}\left(E\right)$ is the contrast material attenuation
coefficient, and $t_{c}$ its thickness. The incident photons were
divided equally between the two spectra.

The K-edge signal $S_{K}$ is the difference of the logarithms of
the photon counts
\begin{equation}
S_{K}=\log\left(N_{K-}\right)-\log\left(N_{K+}\right).\label{eq:K-edge-Signal}
\end{equation}

\noindent{}Using Equations \ref{eq:nKs}, this signal is
\begin{equation}
\begin{aligned}S_{K} &  & = & \left[\mu_{s}\left(E_{k+}\right)-\mu_{s}\left(E_{k+}\right)\right](t_{s}-t_{c})+\ldots\\
 &  &  & \left[\mu_{c}\left(E_{k+}\right)-\mu_{c}\left(E_{k+}\right)\right]t_{c}.
\end{aligned}
\label{eq:Kedge-signal}
\end{equation}

\noindent{}In the background region, the contrast material thickness
is zero, $t_{c}=0$, and the background material thickness is $t_{s}=T_{s}$
so the signal is
\begin{equation}
S_{Kb}=T_{s}\left[\mu_{s}\left(E_{k+}\right)-\mu_{s}\left(E_{k-}\right)\right].\label{eq:K-back-signal}
\end{equation}

\noindent{}Since the soft tissue attenuation coefficient function
$\mu_{s}(E)$ is continuous, the K-edge subtraction background region
signal can be made arbitrarily small by using measurement energies
sufficiently close to the K-edge energy $E_{k}$
\begin{equation}
\begin{array}{ccc}
E_{k-},E_{k+} & \rightarrow & E_{k}\\
\mu_{s}\left(E_{k-}\right) & \rightarrow & \mu_{s}\left(E_{k+}\right)\\
S_{Kb} & \rightarrow & 0
\end{array}\label{eq:K-edge-sig-soft}
\end{equation}
In the contrast material region, the background material thickness
is $T_{s}-T_{c}$ and the contrast thickness is $T_{c}$ so the K-edge
signal from Eq. \ref{eq:Kedge-signal} is

\begin{equation}
\begin{aligned}S_{Kc} &  & = & \left[\mu_{s}\left(E_{k+}\right)-\mu_{s}\left(E_{k+}\right)\right](T_{s}-T_{c})+\ldots\\
 &  &  & \left[\mu_{c}\left(E_{k+}\right)-\mu_{c}\left(E_{k+}\right)\right]T_{c}.
\end{aligned}
\label{eq:S-contrast}
\end{equation}

\noindent{}For the ideal energies specified in Eq. \ref{eq:K-edge-sig-soft},
the first term is essentially zero so the signal in the contrast region
is
\begin{equation}
S_{Kc}=\left[\mu_{c}\left(E_{k+}\right)-\mu_{c}\left(E_{k+}\right)\right]T_{c}.\label{eq:K-contrast-signal}
\end{equation}

The probability distributions of the photon counts in Equations \ref{eq:nKs}
can be modeled as independent Poisson since they are measured with
different spectra. The variance of the logarithm of a Poisson random
variable with expected value $\left\langle n\right\rangle $ is $\nicefrac{1}{\left\langle n\right\rangle }$\cite{Alvarez2010}.
Therefore, the variance of the K-edge signal, which is the difference
of their logarithms, Eq. \ref{eq:Kedge-signal}, is the sum of the
variances of the logarithms of the individual counts 
\begin{equation}
\mathit{Var}\left(S_{K}\right)=\frac{1}{\left\langle N_{K-}\right\rangle }+\frac{1}{\left\langle N_{K+}\right\rangle }.\label{eq:Var-SK}
\end{equation}

\noindent{}The expected values are sufficiently large that we can
use the normal approximation to the Poisson\cite{alvarez2013dimensionality}.

\subsection{A-space processing\label{sub:A-space-processing}}

The A-space method\cite{Alvarez1976} approximates the attenuation
coefficient at points within the object as a linear combination of
basis functions of energy multiplied by coefficients that depend on
the position $\mathbf{r}$ within the object. 
\begin{equation}
\mu(\mathbf{r},E)=a_{1}(\mathbf{r)}\mu_{1}(E)+a_{2}(\mathbf{r})\mu_{2}(E).\label{eq:mu-a1-f1}
\end{equation}
\noindent{}To apply this method to the imaging task in Section \ref{sub:The-imaging-task},
we use the attenuation coefficients of the soft tissue and contrast
material, $\mu_{s}(E)$ and $\mu_{c}(E)$, as the basis functions\cite{Alvarez1979,Alvarez2010}.
With this basis set, the vectors of the basis set coefficients of
the soft tissue and contrast materials are $\mathbf{a_{s}}=[1\ 0]^{T}$
and $\mathbf{a_{c}}=[0\ 1]^{T}$. The superscript $T$ denotes a transpose.

In general, we need a three function basis set to approximate the
attenuation coefficients of biological materials and a high atomic
number contrast agent accurately\cite{alvarez2013dimensionality}.
However, the two function set is sufficient to represent the materials
in the object, which is assumed to be composed only of two materials,
and it facilitates the comparison with the K-edge method. 

The line integral of the attenuation coefficient along a line $\mathcal{L}$
from the source to a detector pixel is 
\begin{equation}
\int_{\mathcal{L}}\mu(\mathbf{r},E)d\mathbf{r}=A_{s}\mu_{s}(E)+A_{c}\mu_{c}(E).\label{eq:line-integral}
\end{equation}
 \noindent{}where $A_{i}=\int_{\mathcal{L}}a_{i}(\mathbf{r})d\mathbf{r},\ i=s,c$
and the superscript $T$ denotes a matrix transpose. We summarize
the line integrals as a vector $\mathbf{A}=[A_{s}\ A_{c}]^{T},$ the
A-vector. 

The A-space method estimates the A-vector\cite{Alvarez1976,Alvarez2011,alvarez_near_optimal_nnet2017}
from measurements of the transmitted spectra. A photon counting detector
with pulse height analysis is used so the counts in each bin are a
different spectrum measurement. Neglecting scatter and pulse pileup,
the expected value of the count in PHA bin $k$ is
\begin{equation}
\left\langle N_{k}(\mathbf{A})\right\rangle =\int\Pi_{k}(E)n_{inc}(E)e^{-A_{s}\mu_{s}(E)-A_{c}\mu_{c}(E)}\label{eq:Nk-x-y}
\end{equation}

\noindent{}where $n_{inc}(E)$ is the incident spectrum and $\Pi_{k}(E)$
is the idealized bin response for bin $k$ , equal to 1 inside the
bin and 0 elsewhere. The measurements can be summarized as a vector
$\mathbf{L(A)}$ with components 
\[
\mathbf{L_{k}(A)}=-\log\left(\frac{N_{k}\left(\mathbf{A}\right)}{\left\langle N_{k}(0)\right\rangle }\right)
\]

\noindent{}where $\left\langle N_{k}(0)\right\rangle $ is the expected
values of the bin count with zero object thickness. The estimator
inverts $\mathbf{L(A)}$ to compute the best estimate of the A-vector,
$\mathbf{\hat{A}}$, given $\mathbf{L}$.

Since the measurements are random quantities, the A-vector estimates
will also be random. If $\mathbf{C_{A}}$ is their covariance, the
signal to noise ratio for the imaging task is
\begin{equation}
SNR^{2}=\mathbf{\delta A^{T}C_{A}^{-1}\delta A}\label{eq:SNR-dA-CA}
\end{equation}

\noindent{}where $\mathbf{\mathbf{\delta A}}$ is the difference
of the A-vectors in the regions with and without contrast material
and the superscript $-1$ denotes the matrix inverse. The optimal
SNR is computed using the Cramèr-Rao lower bound (CRLB), $\mathbf{C_{A.CRLB}}$,
the minimum covariance for any unbiased estimator\cite{KayV1Chapter3}.
For the number of photons required with material selective imaging,
the CRLB is\cite{alvarez2013dimensionality}
\begin{equation}
\mathbf{C_{A.CRLB}=\left(M^{T}C_{L}^{-1}M\right)}^{-1}.\label{eq:CLRB}
\end{equation}

\noindent{}In this equation, $\mathbf{M=\nicefrac{\partial L}{\partial A}}$
and $\mathbf{C_{L}}$ is the covariance of the $\mathbf{L}$ measurements.
With the assumptions of no pileup and a quantum noise limited detector,
$\mathbf{C_{L}}$ is a diagonal matrix with elements $C_{L,kk}=\nicefrac{1}{\left\langle N_{k}\right\rangle }$
and the matrix $\mathbf{M}$ has elements\cite{Alvarez2010}
\[
M_{ij}=\frac{\partial L_{i}}{\partial A_{j}}=\left\langle \mu_{j}(E)\right\rangle _{\hat{n}_{i}(E)}.
\]
\noindent{}That is, each element is the effective value of basis
function $\mu_{j}(E)$ in the normalized spectrum $\hat{n}_{i}(E)$
\[
\hat{n}_{i}(E)=\frac{\Pi_{i}(E)n_{inc}(E)e^{-A_{s}\mu_{s}(E)-A_{c}\mu_{c}(E)}}{\int\Pi_{i}(E)n_{inc}(E)e^{-A_{s}\mu_{s}(E)-A_{c}\mu_{c}(E)}}.
\]

Because we use the attenuation coefficients of the object materials
as the basis functions, the A-vector is 
\begin{equation}
\mathbf{A}=t_{s}\mathbf{a_{s}}+t_{c}\mathbf{a_{c}}\label{eq:A-vec-general}
\end{equation}

\noindent{}where $t_{s}$ and $t_{c}$ are the thicknesses of the
soft tissue and contrast materials. In the soft tissue only region,
\begin{equation}
\mathbf{A_{soft\ tissue}}=\left[\begin{array}{c}
T_{s}\\
0
\end{array}\right].\label{eq:A1}
\end{equation}
 \noindent{} and in the region with contrast agent 
\begin{equation}
\mathbf{A_{contrast}}=\left[\begin{array}{c}
T_{s}-T_{c}\\
T_{c}
\end{array}\right]\label{eq:A2}
\end{equation}

\noindent{}so the difference vector is 
\[
\mathbf{\delta A=A_{contrast}-A_{no\ contrast}=}T_{c}\left[\begin{array}{c}
-1\\
1
\end{array}\right].
\]
\noindent{}Using Eq. \ref{eq:SNR-dA-CA}, the optimal SNR with the
A-space method is 
\[
SNR_{opt}^{2}=T_{c}^{2}\left[1,\ -1\right]\mathbf{M^{T}C_{L}^{-1}M}\left[\begin{array}{c}
-1\\
1
\end{array}\right].
\]

\subsection{Absorbed energy}

The spectra for the K-edge subtraction and A-space processing are
different so in order to compare the two methods on an equal basis
we need a way to normalize their SNR. The method used is to divide
the SNR by the x-ray energy absorbed in the object, which is used
as a proxy for the x-ray dose. The absorbed energy is computed from
the energy absorption coefficient of the soft tissue material, $\mu_{abs,s}(E)$,
which measures the energy absorbed by the object from the incident
x-ray photons\cite{NISTegyAbsorbCoeff}. Assuming the contrast material
is so thin that it does not absorb significantly, the absorbed energy
is 
\begin{equation}
Q_{abs}=Q_{inc}\left[1-\int\hat{q}_{inc}(E)e^{-T_{s}\mu_{abs,s}(E)}dE\right].\label{eq:absorbed-egy}
\end{equation}
\noindent{}In this equation, the incident energy spectrum is $q_{inc}(E)=En_{inc}(E)$,
where $n_{inc}(E)$ is the photon number spectrum incident on the
object. The energy spectrum $q_{inc}(E)dE$ gives the sum of energies
of the photons from $E$ to $E+dE$. The normalized energy spectrum
is
\[
\hat{q}_{inc}(E)=\frac{q_{inc}(E)}{\int q_{inc}(E)dE}.
\]
\noindent{}where the denominator is the total energy of the incident
photons
\[
Q_{inc}=\int q_{inc}(E)dE.
\]

For the A-space method, the absorbed energy is given by Eq. \ref{eq:absorbed-egy}
with the incident x-ray tube spectrum $n_{inc}(E)$ calculated with
the TASMIP algorithm\cite{Boone1997}.

For the idealized K-edge method described in Section \ref{sub:Ideal-K-edge-subtraction},
the two delta function spectra are assumed to be arbitrarily close
to the K-edge energy, $E_{K}$, with a total number of photons for
both spectra equal to $N_{0}$. With these assumptions, the absorbed
energy is
\begin{equation}
Q_{abs,K}=N_{0}E_{K}\left[1-e^{-T_{s}\mu_{abs,s}(E_{K})}\right].\label{eq:Qabsorbed-K}
\end{equation}

\subsection{SNR per absorbed energy as a function of object thickness}

To compare the methods, the $SNR^{2}$ divided by the absorbed energy
was computed as a function of the soft-tissue material thickness from
5 to 25 $g/cm^{2}$. Contrast agents with iodine or gadolinium as
the high atomic number element were used. The contrast material thickness
was fixed at $5\times10^{-3}$ $g/cm^{2}$. The absorbed energies
with the two methods was computed as described in Section \ref{sub:The-imaging-task}. 

For the K-edge method, delta function spectra with energies infinitesimally
below and above the contrast material's K-edge energy were assumed
as described in Section \ref{sub:Ideal-K-edge-subtraction}. The total
incident photons were equally distributed between the two spectra. 

For the A-space method, a 100 kV x-ray tube spectrum computed with
the TASMIP algorithm\cite{Boone1997} was used with photon counting
detectors with five PHA bins. The PHA bins were adjusted to give an
equal number of photons per bin for the spectrum transmitted through
$5$ $g/cm^{2}$ of soft tissue. An efficient estimator with low bias
and with A-vector noise covariance approximately equal to the CRLB
was assumed.

\subsection{SNR per absorbed energy as a function of tube voltage}

The spectrum used with the A-space processing is controlled by the
x-ray tube voltage so the SNR per dose was computed as a function
of voltages from 40 to 100 kV. The object thickness was 20 $g/cm^{2}$.
For comparison, the SNR per dose of the K-edge method with this object
thickness was also plotted.

\section{RESULTS}

\subsection{SNR vs. object thickness}

\begin{figure*}[t]
\centering{}(a)\includegraphics[scale=0.65]{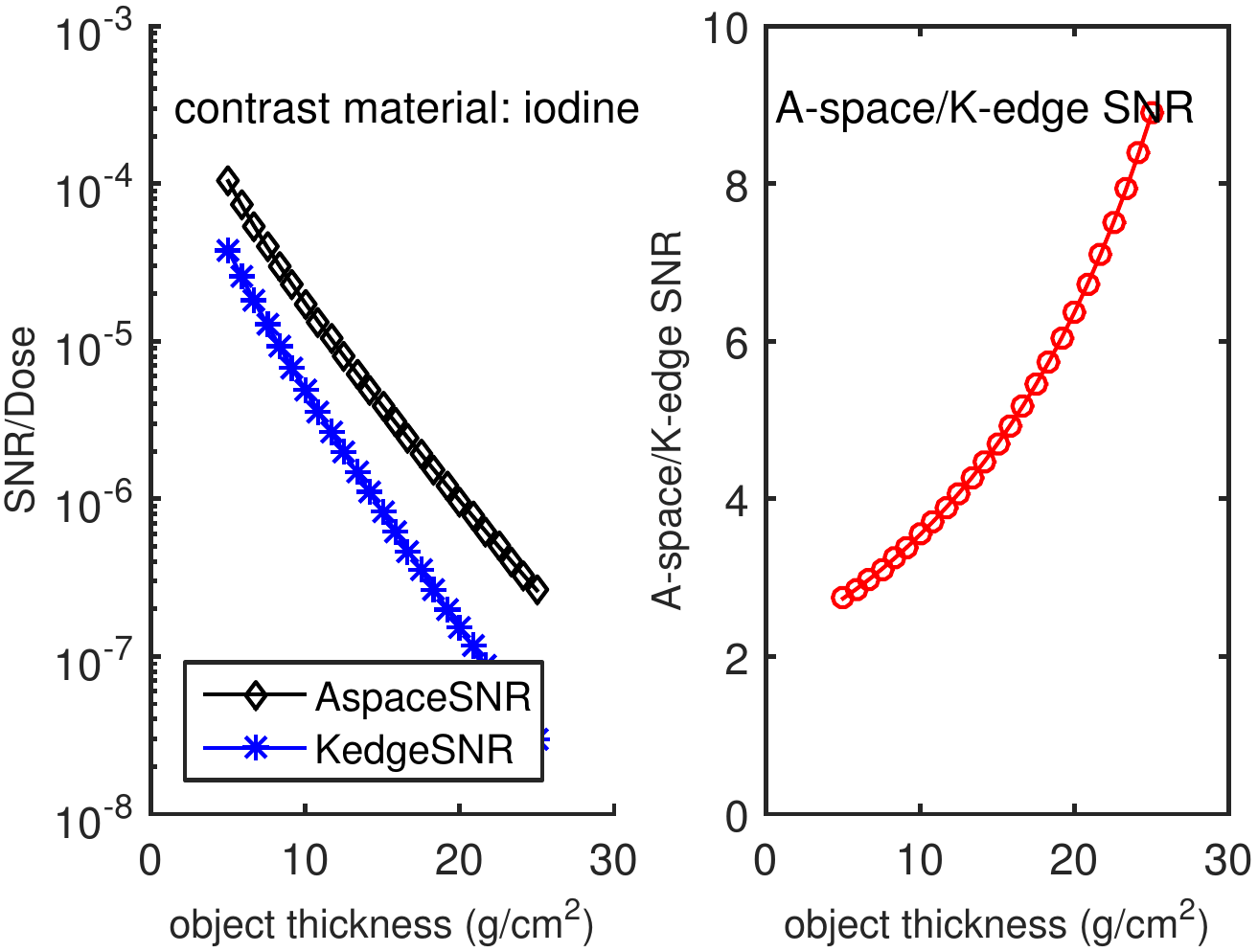}

\centering{}(b)\includegraphics[scale=0.65]{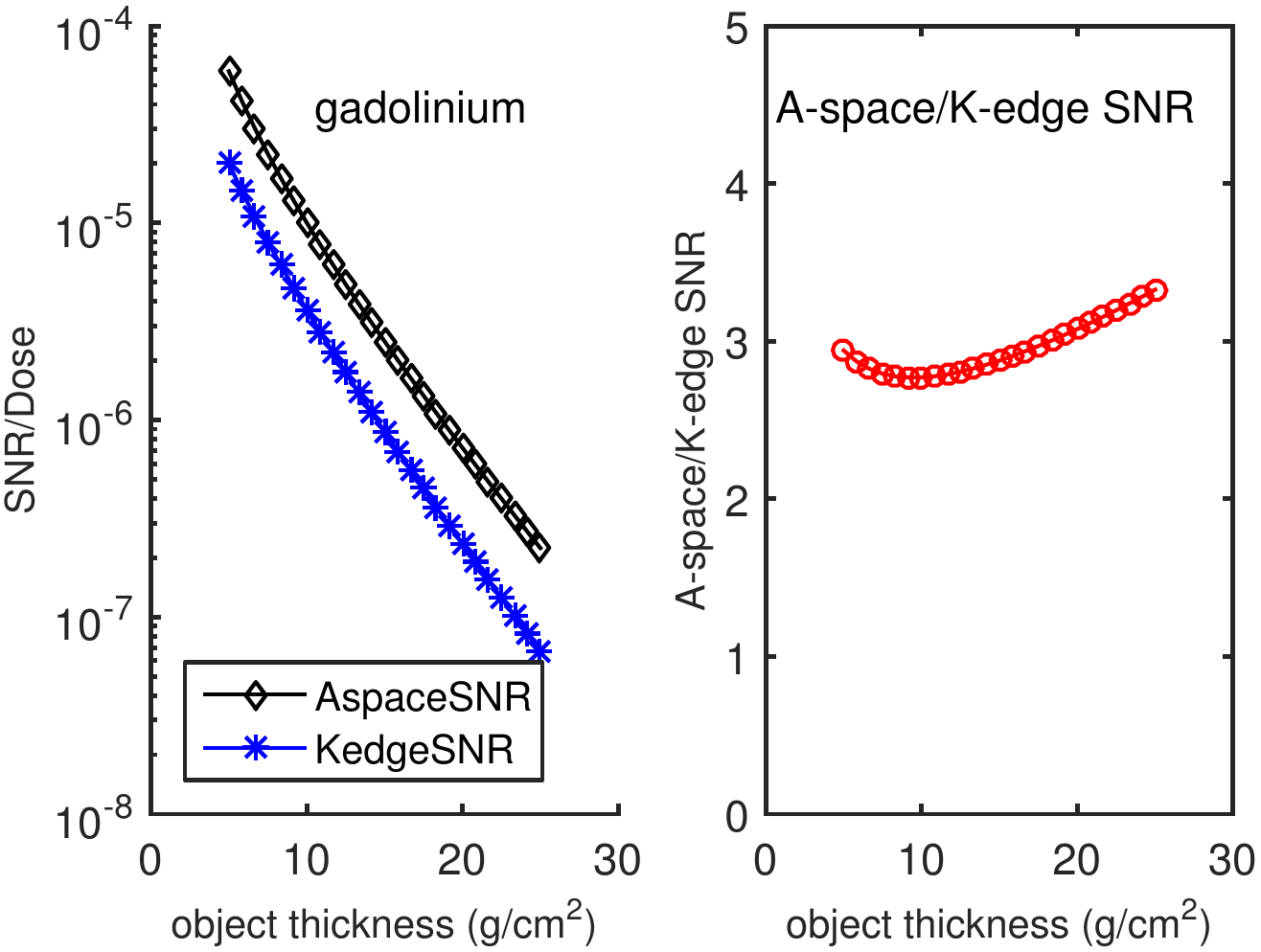}

\protect\caption{Contrast agent SNR per unit dose for K-edge and A-space methods as
a function of object thickness for iodine (a) and gadolinium (b) containing
contrast agents. For each material, the left panels shows the SNR
per unit dose while the right panel shows the ratio of the SNR/dose.
The tube voltage for the A-space method was 100 kV. \label{fig:SNR-thickness}}
\end{figure*}

Figure \ref{fig:SNR-thickness} shows the SNR per dose as a function
of the soft tissue thickness. Panel (a) is for an iodine containing
contrast agent and Panel (b) for gadolinium containing contrast agent
. In the panels, the SNR of the two methods are on the left and their
ratio on the right. The tube voltage for the A-space method was 100
kV.

\subsection{SNR vs. tube voltage}

\begin{figure*}
\centering{}\includegraphics[scale=0.55]{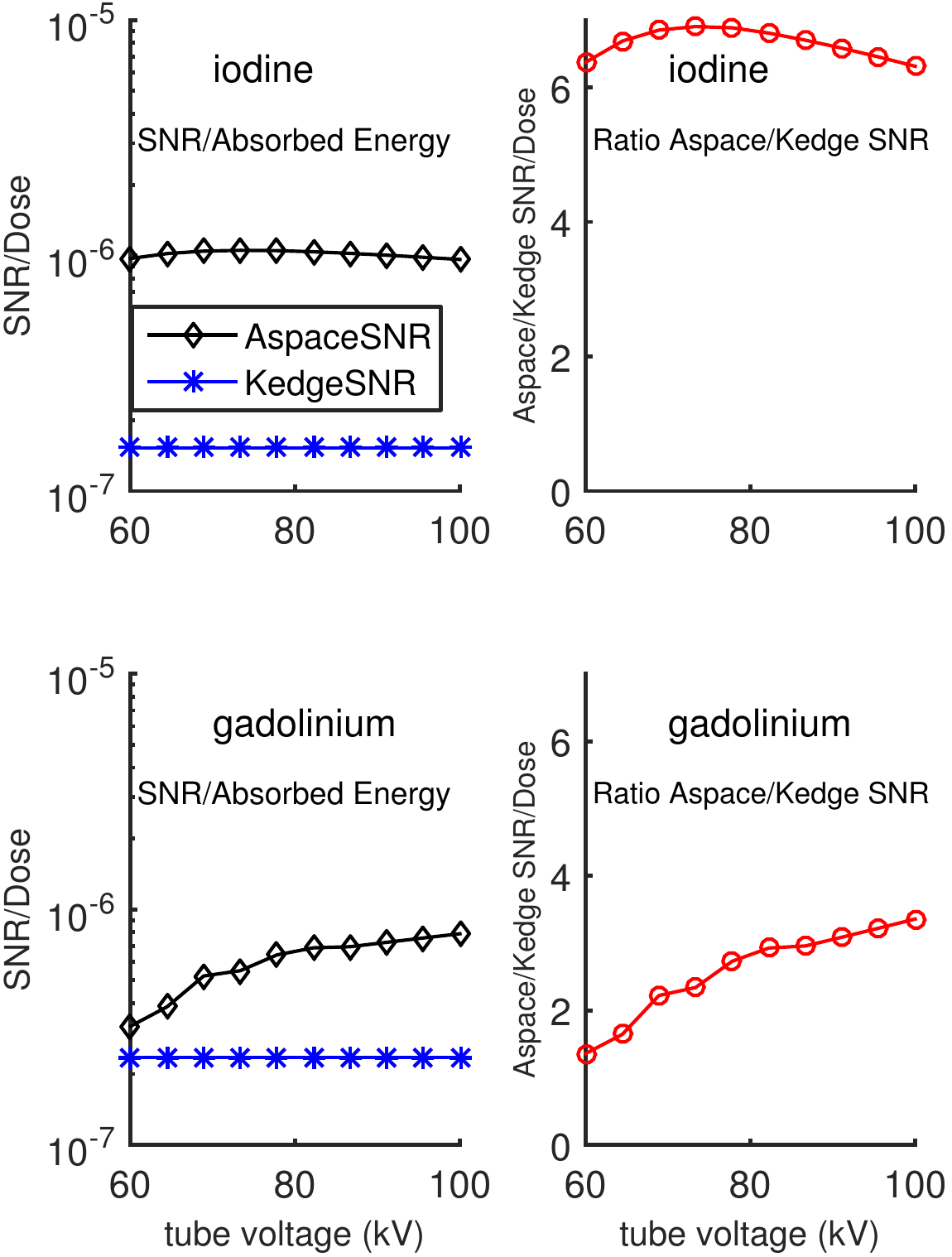}

\protect\caption{SNR per absorbed energy versus tube voltage. Iodine contrast in the
top and gadolinium in the bottom panels. The K-edge method result
does not depend on the tube spectrum so its SNR is constant. The panels
on the right show the ratios of the SNRs. The object thickness was
fixed at 20 $g/cm^{2}$. \label{fig:SNR-kV}}
\end{figure*}

Figure \ref{fig:SNR-kV} shows the SNR per absorbed energy as a function
of the x-ray tube voltage. The left graph is for an iodine contrast
agent and the right for gadolinium contrast. The K-edge method SNR
per dose is also plotted. Since it does not depend on the tube spectrum,
it is a constant. The object thickness was fixed at 20 $g/cm^{2}$.

\section{DISCUSSION}

Figure \ref{fig:SNR-thickness} shows that for soft tissue thickness
from 5 to 25 $g/cm^{2}$ the A-space method's SNR per dose is larger
than that of the K-edge subtraction method. The ratio of the SNR values
with the two methods depends on the contrast agent. For iodine containing
agents, the ratio becomes larger as the object thickness increases
approaching a value of 9 at 25 $g/cm^{2}$. The gadolinium ratio is
approximately constant with a value approximately 3. The K-edge energy
of iodine, 33.2 keV, is substantially lower than that of gadolinium,
50.2 keV, so the iodine K-edge signal is attenuated much more strongly
as the object thickness increases. 

Figure \ref{fig:SNR-kV} shows that for tube voltages greater than
60 kV, the A-space method SNR is larger than the K-edge subtraction
method SNR for both iodine and gadolinium. For iodine, the A-space
method SNR is approximately 6 times larger than the K-edge subtraction
value. For gadolinium, the A-space SNR is close to the K-edge subtraction
value at low tube voltages but increases to approximately 3.5 times
larger at 100 kV. The change may be due to the fact that for low voltages
the peak of the tube spectrum is close to the K-edge energy while
for larger voltages the signal includes the contributions of the difference
in attenuation of soft tissue and gadolinium above the K-edge energy. 

The larger SNR of the A-space method may seem surprising since the
sharp discontinuity of the contrast material attenuation coefficient
at the K-edge is markedly different from the background material attenuation
coefficient. The superior SNR of the A-space method is due to the
fact that the contrast material attenuation is different from the
soft tissue not only at the K-edge energy but also throughout the
range of energies in the x-ray tube spectrum. The A-space method measures
this difference throughout the energy region leading to larger signal.
Also, the tube spectrum above the K-edge is attenuated less than at
the K-edge leading to a lower noise measurement. 

The K-edge method implementation in this study used idealized measurement
spectra with high flux and energies tunable infinitesimally close
to the K-edge energy. No sources with these characteristics that can
be deployed to clinical institutions currently exist. Early work\cite{jacobson_dichromatic_1953}
used a broad spectrum x-ray tube source with a crystal monochromator
but this did not provide a practical photon flux suitable for a clinical
system. Other early work\cite{riederer_selective_iodine_Kedge_1977}
used an x-ray tube filtered by appropriately chosen materials. With
this approach, there is a trade-off between the tube loading and the
width of the measurement spectra. For practical tube loading, the
filtered spectra widths are not sufficiently small to give a large
signal across the K-edge discontinuity. Another possibility\cite{dilmanian_synchrotron_Kedge1997}
is a synchrotron radiation source but currently this requires a large
nuclear physics accelerator laboratory. There is research in ``table-top''
synchrotron radiation sources but these have not proved practical
at this time. As shown by the results in this paper, even with an
ideal monoenergetic source, the A-space method provides a better signal
to noise ratio per dose. 

The A-space method used in this study, although idealized, may be
implementable in a clinical environment. X-ray tubes are widely used
in diagnostic imaging. The photon counting detector assumed negligible
pileup and perfect PHA bins but realistic pileup and overlap between
PHA bin responses may not substantially reduce the A-space method
performance\cite{taguchi2013vision,AlvarezSNRwithPileup2014}. The
effect of photon counting detector imperfections on the SNR is a subject
of current research.

\section{CONCLUSION}

The K-edge subtraction and the A-space method for imaging contrast
agents containing iodine or gadolinium in a soft tissue background
material are compared for their signal to noise ratio per unit dose.
The A-space method has a better $SNR^{2}/dose$ for soft tissue object
thicknesses from 10 to 25 $g/cm^{2}$ and for tube voltages above
60 kV.

\section{Supplementary material}

Matlab language code to reproduce the figures of this paper is available
online\cite{alvarez_kedge_paper_code1}.



\begin{thebibliography}{10}
\providecommand{\url}[1]{#1}
\csname url@samestyle\endcsname
\providecommand{\newblock}{\relax}
\providecommand{\bibinfo}[2]{#2}
\providecommand{\BIBentrySTDinterwordspacing}{\spaceskip=0pt\relax}
\providecommand{\BIBentryALTinterwordstretchfactor}{4}
\providecommand{\BIBentryALTinterwordspacing}{\spaceskip=\fontdimen2\font plus
\BIBentryALTinterwordstretchfactor\fontdimen3\font minus
  \fontdimen4\font\relax}
\providecommand{\BIBforeignlanguage}[2]{{%
\expandafter\ifx\csname l@#1\endcsname\relax
\typeout{** WARNING: IEEEtran.bst: No hyphenation pattern has been}%
\typeout{** loaded for the language `#1'. Using the pattern for}%
\typeout{** the default language instead.}%
\else
\language=\csname l@#1\endcsname
\fi
#2}}
\providecommand{\BIBdecl}{\relax}
\BIBdecl

\bibitem{alvarez_kedge_paper_code1}
\BIBentryALTinterwordspacing
R.~Alvarez, ``Matlab language software for paper: "{K}-edge subtraction vs.
  {A}-space processing for x-ray imaging of contrast agents: {SNR}",'' July
  2017. [Online]: \url{http://dx.doi.org/ 10.13140/RG.2.2.30400.64009}
\BIBentrySTDinterwordspacing

\bibitem{jacobson_dichromatic_1953}
\BIBentryALTinterwordspacing
B.~Jacobson, ``\BIBforeignlanguage{en}{Dichromatic {Absorption} {Radiography}.
  {Dichromography}},'' \emph{\BIBforeignlanguage{en}{Acta Radiologica}}, vol.
  Original Series, Volume 39, no.~6, pp. 437--452, Jun. 1953. [Online]:
  \url{http://tinyurl.com/JacobsonDichromatic1953}
\BIBentrySTDinterwordspacing

\bibitem{jacobson_edholm_iodine_1959}
\BIBentryALTinterwordspacing
P.~Edholm and B.~Jacobson, ``\BIBforeignlanguage{en}{Quantitative
  {Determination} of {Iodine} in {Vivo}},'' \emph{\BIBforeignlanguage{en}{Acta
  Radiologica}}, vol. Original Series, Volume 52, no.~5, pp. 337--346, Nov.
  1959. [Online]: \url{http://tinyurl.com/EdholmJacobson1959}
\BIBentrySTDinterwordspacing

\bibitem{riederer_selective_iodine_Kedge_1977}
\BIBentryALTinterwordspacing
S.~J. Riederer and C.~A. Mistretta, ``\BIBforeignlanguage{en}{Selective iodine
  imaging using k-edge energies in computerized x-ray tomography},''
  \emph{\BIBforeignlanguage{en}{Medical Physics}}, vol.~4, no.~6, pp. 474--481,
  Nov. 1977. [Online]: \url{http://dx.doi.org/10.1118/1.594357}
\BIBentrySTDinterwordspacing

\bibitem{Alvarez1976}
\BIBentryALTinterwordspacing
R.~E. Alvarez and A.~Macovski, ``{Energy-selective reconstructions in X-ray
  computerized tomography},'' \emph{Phys. Med. Biol.}, vol.~21, pp. 733--44,
  1976. [Online]: \url{http://dx.doi.org/10.1088/0031-9155/21/5/002}
\BIBentrySTDinterwordspacing

\bibitem{Alvarez2010}
\BIBentryALTinterwordspacing
R.~E. Alvarez, ``{Near optimal energy selective x-ray imaging system
  performance with simple detectors},'' \emph{Med. Phys.}, vol.~37, pp.
  822--841, 2010. [Online]: \url{http://tinyurl.com/NearOptimalEnergySelective}
\BIBentrySTDinterwordspacing

\bibitem{Tapiovaara1985}
\BIBentryALTinterwordspacing
M.~J. Tapiovaara and R.~Wagner, ``{SNR and DQE analysis of broad spectrum X-ray
  imaging},'' \emph{Phys. Med. Biol.}, vol.~30, pp. 519--529, 1985. [Online]:
  \url{http://doi.org/10.1088/0031-9155/30/6/002}
\BIBentrySTDinterwordspacing

\bibitem{Alvarez}
\BIBentryALTinterwordspacing
R.~E. Alvarez, ``{Extraction of Energy Dependent Information in Radiography},''
  Ph.D. dissertation, Stanford University, 1976. [Online]:
  \url{http://www.dx.doi.org/10.13140/RG.2.2.12965.09446}
\BIBentrySTDinterwordspacing

\bibitem{Alvarez1982}
\BIBentryALTinterwordspacing
------, ``Energy dependent information in x-ray imaging---part 1 the vector
  space description,'' \emph{Stanford University Information Systems
  Laboratory-unpublished}, 1982. [Online]:
  \url{http://dx.doi.org/10.13140/RG.2.2.19355.46887}
\BIBentrySTDinterwordspacing

\bibitem{Lehmann1981}
L.~A. Lehmann, R.~E. Alvarez, A.~Macovski, W.~R. Brody, N.~J. Pelc, S.~J.
  Riederer, and A.~L. Hall, ``{Generalized image combinations in dual KVP
  digital radiography},'' \emph{Med. Phys.}, vol.~8, pp. 659--67, 1981.

\bibitem{Roessl2007}
\BIBentryALTinterwordspacing
E.~Roessl and R.~Proksa, ``{K-edge imaging in x-ray computed tomography using
  multi-bin photon counting detectors},'' \emph{Phys. Med. Biol.}, vol.~52, pp.
  4679--4696, 2007. [Online]:
  \url{http://dx.doi.org/10.1088/0031-9155/52/15/020}
\BIBentrySTDinterwordspacing

\bibitem{ZimmermanSPIEMedImg3Destimators2016}
\BIBentryALTinterwordspacing
K.~C. Zimmerman and T.~Gilat~Schmidt, ``Comparison of quantitative {K}-edge
  empirical estimators using an energy-resolved photon-counting detector,''
  D.~Kontos, T.~G. Flohr, and J.~Y. Lo, Eds., Mar. 2016, p. 97831S. [Online]:
  \url{http://dx.doi.org/10.1117/12.2217233}
\BIBentrySTDinterwordspacing

\bibitem{dilmanian_synchrotron_Kedge1997}
\BIBentryALTinterwordspacing
F.~A. Dilmanian, X.~Y. Wu, E.~C. Parsons, B.~Ren, J.~Kress, T.~M. Button, L.~D.
  Chapman, J.~A. Coderre, F.~Giron, D.~Greenberg, D.~J. Krus, Z.~Liang,
  S.~Marcovici, M.~J. Petersen, C.~T. Roque, M.~Shleifer, D.~N. Slatkin, W.~C.
  Thomlinson, K.~Yamamoto, and Z.~Zhong, ``Single- and dual-energy {CT} with
  monochromatic synchrotron x-rays,'' \emph{Phys. Med. Biol.}, vol.~42, no.~2,
  pp. 371--387, Feb. 1997. [Online]:
  \url{http://www.dx.doi.org/10.1088/0031-9155/42/2/009}
\BIBentrySTDinterwordspacing

\bibitem{shikhaliev2012photonCountKedgeFiltered}
\BIBentryALTinterwordspacing
P.~M. Shikhaliev, ``Photon counting spectral {CT}: improved material
  decomposition with {K}-edge-filtered x-rays,'' \emph{Phys. Med Biol.},
  vol.~57, no.~6, p. 1595, 2012. [Online]:
  \url{http://www.dx.doi.org/10.1088/0031-9155/57/6/1595}
\BIBentrySTDinterwordspacing

\bibitem{VanTrees2013}
\BIBentryALTinterwordspacing
H.~Van~Trees, K.~Bell, and Z.~Tian, \emph{Detection Estimation and Modulation
  Theory, Detection, Estimation, and Filtering Theory}, ser. Detection
  Estimation and Modulation Theory.\hskip 1em plus 0.5em minus 0.4em\relax New
  York: Wiley, 2013. [Online]:
  \url{https://books.google.com/books?id=dnvaxqHDkbQC}
\BIBentrySTDinterwordspacing

\bibitem{alvarez2013dimensionality}
\BIBentryALTinterwordspacing
R.~E. Alvarez, ``Dimensionality and noise in energy selective x-ray imaging,''
  \emph{Med. Phys.}, vol.~40, no.~11, p. 111909, 2013. [Online]:
  \url{http://dx.doi.org/10.1118/1.4824057}
\BIBentrySTDinterwordspacing

\bibitem{Alvarez1979}
\BIBentryALTinterwordspacing
R.~E. Alvarez and E.~J. Seppi, ``{A comparison of noise and dose in
  conventional and energy selective computed tomography},'' \emph{{IEEE} Trans.
  Nucl. Sci.}, vol. NS-26, pp. 2853--2856, 1979. [Online]:
  \url{http://www.dx.doi.org/10.1109/TNS.1979.4330549}
\BIBentrySTDinterwordspacing

\bibitem{Alvarez2011}
\BIBentryALTinterwordspacing
R.~E. Alvarez, ``{Estimator for photon counting energy selective x-ray imaging
  with multi-bin pulse height analysis},'' \emph{Med. Phys.}, vol.~38, pp.
  2324--2334, 2011. [Online]: \url{http://dx.doi.org/10.1118/1.3570658}
\BIBentrySTDinterwordspacing

\bibitem{alvarez_near_optimal_nnet2017}
\BIBentryALTinterwordspacing
R.~Alvarez, ``Near optimal neural network estimator for spectral x-ray photon
  counting data with pileup,'' \emph{arXiv:1702.01006 [physics]}, Feb. 2017,
  arXiv: 1702.01006. [Online]: \url{http://arxiv.org/abs/1702.01006}
\BIBentrySTDinterwordspacing

\bibitem{KayV1Chapter3}
S.~M. Kay, \emph{{Fundamentals of Statistical Signal Processing, Volume I:
  Estimation Theory }}.\hskip 1em plus 0.5em minus 0.4em\relax Upper Saddle
  River, NJ: Prentice Hall PTR, 1993, vol. Ch. 3.

\bibitem{NISTegyAbsorbCoeff}
\BIBentryALTinterwordspacing
``{NIST}: {X}-{Ray} {Mass} {Attenuation} {Coefficients} - {Section} 3.''
  [Online]: \url{https://physics.nist.gov/PhysRefData/XrayMassCoef/chap3.html}
\BIBentrySTDinterwordspacing

\bibitem{Boone1997}
\BIBentryALTinterwordspacing
J.~M. Boone and J.~A. Seibert, ``{An accurate method for computer-generating
  tungsten anode x-ray spectra from 30 to 140 kV},'' \emph{Med. Phys.},
  vol.~24, pp. 1661--70, 1997. [Online]:
  \url{http://dx.doi.org/10.1118/1.597953}
\BIBentrySTDinterwordspacing

\bibitem{taguchi2013vision}
\BIBentryALTinterwordspacing
K.~Taguchi and J.~S. Iwanczyk, ``Vision 20/20: Single photon counting x-ray
  detectors in medical imaging,'' \emph{Med. Phys.}, vol.~40, p. 100901, 2013.
  [Online]: \url{http://dx.doi.org/10.1118/1.4820371}
\BIBentrySTDinterwordspacing

\bibitem{AlvarezSNRwithPileup2014}
\BIBentryALTinterwordspacing
R.~E. Alvarez, ``Signal to noise ratio of energy selective x-ray photon
  counting systems with pileup,'' \emph{Med. Phys.}, vol.~41, no.~11, p.
  111909, 2014. [Online]: \url{http://dx.doi.org/10.1118/1.4898102}
\BIBentrySTDinterwordspacing

\end{thebibliography}
\end{document}